%% file: main.tex
\documentclass[letterpaper,twocolumn,10pt]{article}
\usepackage{usenix2019_v3}

\usepackage{amsmath,amssymb,amsfonts}
\usepackage{graphicx}
\usepackage{booktabs}
\usepackage{float}
\usepackage{listings}
\usepackage{subcaption}
\usepackage{xspace}
\usepackage{tikz}
\usetikzlibrary{arrows.meta, positioning, calc, shapes.geometric, fit, shadows}

\AtBeginDocument{\DeclareMathAlphabet{\mathcal}{OMS}{cmsy}{m}{n}}

\newcommand{\openkedge}{\mbox{OpenKedge}\xspace}
\newcommand{\sab}{\mbox{SAB}\xspace}
\newcommand{\seb}{\mbox{SEB}\xspace}
\newcommand{\sqa}{\mbox{SQA}\xspace}

\definecolor{slate}{RGB}{112,128,144}
\definecolor{emerald}{RGB}{80,200,120}

\begin{document}

\title{\bf Sovereign Execution Broker: Enforcing Certificate-Bound Authority in Agentic Control Planes}

\author{
  {\rm Jun He}\\
  OpenKedge.io
  \and
  {\rm Deying Yu}\\
  OpenKedge.io
}

\maketitle

\input{sections/01-introduction}
\input{sections/02-background-motivation}
\input{sections/03-system-threat-model}
\input{sections/04-execution-model}
\input{sections/05-scoped-identity}
\input{sections/06-drift-revocation-toctou}
\input{sections/07-implementation}
\input{sections/08-evaluation}
\input{sections/09-security-analysis}
\input{sections/10-discussion-limitations}
\input{sections/11-related-work}
\input{sections/12-conclusion}

\bibliographystyle{unsrt}
\bibliography{refs}

\end{document}

%% file: sections/01-introduction.tex
\begin{abstract}
Autonomous agents are increasingly connected to cloud,
deployment, and data-control workflows, but production
mutation authority should not reside inside non-deterministic
reasoning processes. Existing access-control mechanisms
authorize identities, while assurance layers certify proposed
actions; neither alone provides a mandatory enforcement point for
certified authority at the moment of mutation. This paper introduces the
Sovereign Execution Broker (SEB), a runtime enforcement
boundary for certificate-bound agentic infrastructure. SEB
consumes certificates issued by the Sovereign Assurance Boundary (SAB),
verifies that
the requested mutation matches the certified execution
contract, checks validity windows, policy epochs, revocation
epochs, and live-state drift, mints scoped execution identity,
invokes infrastructure APIs, and records signed decision and outcome
records.
By separating proposal, admission, and execution, SEB
turns certified authority into a short-lived, revocable, auditable
runtime capability, provided that production mutation APIs reject
non-broker identities. We present the SEB execution model,
certificate and replay-verification predicates, scoped identity
semantics, bypass-prevention deployment patterns, failure behavior,
and a concrete prototype implementation. We evaluate the prototype
on AWS and Kubernetes clusters, measuring latency overheads,
revocation propagation, drift detection, and security under fault injection.
\end{abstract}

\section{Introduction}
\label{sec:intro}

As autonomous systems powered by Large Language Models (LLMs) transition from passive analytical assistants to active operational planners, they are increasingly integrated into production infrastructure. Today, autonomous agents are tasked with auto-scaling compute groups, deploying containers, configuring cloud security groups, and managing databases. However, because these agents use non-deterministic reasoning processes, granting them direct access credentials creates severe security and operational risks. A single hallucination or adversarial prompt injection can trigger unauthorized or destructive infrastructure mutations.

To mitigate these risks, recent architectures have introduced institutional admission gates such as the Sovereign Assurance Boundary (\sab)~\cite{he2026sovereignassurance} to certify agentic proposals before they are executed. \sab evaluates proposed actions against safety and operational contracts, producing a cryptographically signed certificate ($\Omega$) upon successful validation. However, an admission certificate is not enforcement unless the target infrastructure mutation paths actively require a broker to verify it. 

In the \openkedge stack, Semantic Quorum Assurance (\sqa) validates proposed actions, \sab certifies admitted authority, and \seb enforces the resulting certificates at mutation time.

If agents or their runtime wrappers continue to hold standing production credentials, the assurance layer can be bypassed entirely. For example, a compromised wrapper or a direct API execution path could allow an agent to perform actions without ever submitting its proposals to the admission gate. Moreover, even when an agent complies with the admission process, the time lag between proposal admission and actual execution creates a window of vulnerability where the underlying state can drift, or the security environment can change.

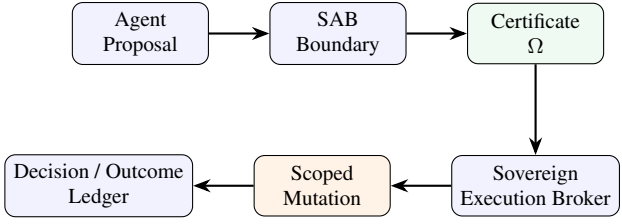
\begin{figure}[t]
  \centering
  \begin{tikzpicture}[
    node distance=1.0cm and 0.8cm,
    block/.style={draw, rectangle, minimum height=0.8cm, minimum width=1.8cm, rounded corners, align=center, fill=blue!5, font=\footnotesize},
    arrow/.style={-Stealth, thick}
  ]
    \node (agent) [block] {Agent\\Proposal};
    \node (airlock) [block, right=of agent] {SAB\\Boundary};
    \node (cert) [block, right=of airlock, fill=emerald!10] {Certificate\\$\Omega$};
    \node (seb) [block, below=of cert, yshift=-0.2cm] {Sovereign\\Execution Broker};
    \node (mut) [block, left=of seb, fill=orange!10] {Scoped\\Mutation};
    \node (ledger) [block, left=of mut] {Decision / Outcome\\Ledger};

    \draw [arrow] (agent) -- (airlock);
    \draw [arrow] (airlock) -- (cert);
    \draw [arrow] (cert) -- (seb);
    \draw [arrow] (seb) -- (mut);
    \draw [arrow] (mut) -- (ledger);
  \end{tikzpicture}
  \caption{\openkedge pipeline from agent proposal to runtime enforcement. In the intended deployment, the Sovereign Execution Broker is the only autonomous-path entity possessing credentials to mutate state, which it does only upon verifying a valid certificate $\Omega$ and writing signed decision or outcome records to the ledger.}
  \label{fig:pipeline}
\end{figure}

To address this gap, we introduce the Sovereign Execution Broker (\seb). As shown in Figure~\ref{fig:pipeline}, \seb is the runtime enforcement boundary for certificate-bound agentic control planes. In the intended deployment, no agent runtime holds standing mutation credentials. Instead, all autonomous mutation requests must go through the \seb, which acts as the custodian of execution authority. At the moment of mutation, \seb verifies the SAB certificate, matches the request against the certified contract, checks revocation and drift, mints short-lived scoped credentials, executes the mutation, and logs signed decision or outcome records.

\paragraph{Contributions.}
This paper makes four contributions:

\begin{enumerate}
    \item \textbf{Broker-enforced autonomy.}
    We identify the enforcement gap between certified autonomous
    proposals and actual infrastructure mutation, and define the
    Sovereign Execution Broker as the runtime boundary that prevents
    autonomous agents from holding or exercising standing production
    credentials.

    \item \textbf{A certificate-verification execution model.}
    We formalize the broker execution interface
    \(Execute(\Omega, req, S_t, Platform) \rightarrow D \mid O\), where the broker
    verifies the SAB certificate \(\Omega\), matches
    the requested mutation against the certified contract \(C\),
    checks validity windows, policy epochs, revocation epochs, and
    live-state drift before executing.

    \item \textbf{Scoped execution identity and revocation-before-execution.}
    We introduce a least-privilege identity model in which credentials
    are minted only at execution time, scoped to the certified contract,
    target resource, validity window, and revocation epoch. This
    ensures that pending certificates can be invalidated before mutation
    and that agents never receive reusable mutation authority.

    \item \textbf{Prototype implementation and empirical evaluation.}
    We implement the broker as a Go service with adapters for AWS STS and
    Kubernetes TokenRequest, and deploy it as a mandatory mutation path.
    We evaluate the prototype under realistic workloads, reporting p50,
    p95, and p99 execution latencies, microbenchmarks for cryptographic and
    drift validation, revocation propagation delays, and behavior under
    induced partition and crash-recovery scenarios.
\end{enumerate}

%% file: sections/02-background-motivation.tex
\section{Background and Motivation}
\label{sec:background}

SEB is motivated by a mismatch between identity-centric access control and certificate-bound autonomous mutations.

\subsection{Why IAM is Insufficient}
Traditional Identity and Access Management (IAM) systems are designed under the assumption of human-driven or static programmatic execution. An IAM policy typically answers a binary authorization query: 
\begin{center}
    \emph{``Is this principal (e.g., identity or role) authorized to perform this API operation on this resource?''}
\end{center}
In agentic control planes, this model fails. Autonomous agents reason dynamically and non-deterministically; thus, granting an agent static IAM credentials means trusting all subsequent logic generated by that agent. If the agent's prompts are hijacked or its reasoning loop fails, those static permissions allow the agent to execute arbitrary destructive operations (e.g., deleting databases or opening public ports). 

In contrast, a Sovereign Execution Broker (\seb) evaluates a much richer context. Rather than validating static identity permissions, \seb answers:
\begin{center}
    \emph{``Is this specific, certified mutation request still valid under the current system evidence, active policy epoch, revocation epoch, live-state drift, and cryptographic scope?''}
\end{center}
By shifting authorization from the identity to the certified action itself, \seb removes standing mutation credentials from agent runtimes when the deployment denies all non-broker mutation identities.

\input{sections/table-mechanisms}

\subsection{Why SAB Needs a Broker}
The Sovereign Assurance Boundary (\sab) provides institutional admission control by validating the semantic safety of proposed agent actions~\cite{he2026sovereignassurance}. At the conclusion of validation, \sab emits a signed certificate $\Omega$. However, $\Omega$ is merely a cryptographic assertion. Without a runtime broker that acts as the sole credential holder, $\Omega$ becomes a post-hoc audit artifact. An agent could easily bypass the admission process by calling APIs directly using standing credentials, or execute expired/invalidated certificates. 

Therefore, \sab and \seb form a complementary pair: \sab decides whether a proposal \emph{may} become authority at admission time, whereas \seb enforces that authority at the moment of mutation.

\subsection{Running Example}
To illustrate the need for \seb, consider a security group change where an autonomous agent $A_{ops}$ is tasked with opening a firewall port for a new service.
\begin{enumerate}
    \item \textbf{Proposal:} $A_{ops}$ proposes a contract $C_{fw}$ to open port $443$ to CIDR $10.0.0.0/16$.
    \item \textbf{Admission:} \sab validates $C_{fw}$ against corporate firewall policy and issues certificate $\Omega_{fw}$.
    \item \textbf{Intervening Event:} Before execution, the security environment drifts: the subnet is decommissioned, or a global emergency revocation epoch advances.
    \item \textbf{Runtime Check:} When the wrapper requests execution, the \seb intercepts the request and checks the live state $S_t$. Because the target subnet is invalid or the certificate's epoch is stale, \seb rejects the execution.
    \item \textbf{Execution:} If all checks pass, \seb mints a short-lived IAM token bound strictly to the target security group and the port-opening operation, executes it, and logs the outcome.
\end{enumerate}
This flow prevents the agent from holding standing credentials to edit the security group, and the broker rejects replay under drifted conditions, assuming the cloud API rejects non-broker mutation identities.

%% file: sections/table-mechanisms.tex
\begin{table}[t]
\centering
\caption{\seb compared to existing authorization and safety enforcement mechanisms.}
\label{tab:mechanisms}
\small
\begin{tabular}{lp{2.7cm}p{2.7cm}}
\toprule
\textbf{Mechanism} & \textbf{Enforces} & \textbf{Missing} \\
\midrule
IAM & Identity permissions & Proposal semantics \\
OPA & Policy predicates & Certificate-bound authority \\
Audit Log & After-the-fact record & Pre-execution prevention \\
SAB & Admission certificate & Mutation-time enforcement \\
\midrule
\bf \seb & \bf Certificate-bound execution & \bf Requires mandatory mutation-path deployment \\
\bottomrule
\end{tabular}
\end{table}

%% file: sections/03-system-threat-model.tex
\section{System and Threat Model}
\label{sec:system-threat}

The SEB threat model separates untrusted autonomous planning from trusted admission, broker execution, and target-platform enforcement.

\subsection{System Components}
The system consists of the following components:
\begin{itemize}
    \item \textbf{Agent Runtime:} An untrusted, non-deterministic reasoning loop (e.g., an LLM-based agent) that plans and proposes infrastructure mutations.
    \item \textbf{SAB Admission Boundary:} The trusted admission gateway that intercepts proposals, verifies their safety using Semantic Quorum Assurance (\sqa) or policy engines, and emits cryptographically signed certificates $\Omega$.
    \item \textbf{Sovereign Execution Broker (SEB):} The trusted execution boundary that intercepts all mutation requests, verifies certificates, checks live state, and performs execution.
    \item \textbf{Target Infrastructure:} The cloud APIs (e.g., AWS, GCP) or orchestrators (e.g., Kubernetes) undergoing mutations.
    \item \textbf{Revocation Service:} A service maintaining the current global policy and revocation epoch $\rho_{active}$.
    \item \textbf{Ledger:} An append-only log capturing the detailed outcomes of broker decisions and executions.
    \item \textbf{Identity Provider / IAM:} The system (e.g., AWS STS, Kubernetes token issuer) that mints credentials.
    \item \textbf{Evidence Sources:} Distributed monitors supplying real-world state evidence to both \sab and \seb.
\end{itemize}

\subsection{Trust Assumptions}
We partition the system components into trusted boundaries and untrusted or fallible components:

\paragraph{Trusted Components.} We assume the integrity of:
\begin{itemize}
    \item The \seb execution environment (e.g., running within a secure enclave or highly restricted private network segment).
    \item The \sab signing key, the \seb signing key, and the corresponding public verification keys.
    \item The global Revocation Epoch Service, which distributes epoch updates.
    \item The local Cloud Provider's Identity Provider (IdP) for minting short-lived session credentials.
    \item The target platform's IAM, RBAC, admission-control, and policy-enforcement mechanisms that reject mutation identities outside the broker path.
    \item The append-only properties of the Ledger.
\end{itemize}

\paragraph{Deployment Assumptions.} The enforcement theorem in
Section~\ref{sec:security-analysis} relies on three operational
configuration assumptions. First, no agent runtime or wrapper possesses
standing credentials that authorize production mutation. Second, we
define \emph{authorized mutation principals} as the broker role plus
broker-issued execution sessions created after certificate verification;
target APIs accept mutation requests only from this set. Third, emergency
break-glass identities are outside the autonomous agent path and are
separately authenticated, authorized, and audited.

\paragraph{Untrusted or Fallible Components.} We do \emph{not} assume the security or correctness of:
\begin{itemize}
    \item The Agent Runtime or its internal prompting context.
    \item Individual semantic validators within the \sqa quorums, which may be compromised or exhibit model bias.
    \item Local operators or administrators (who could attempt malicious emergency overrides).
    \item Live target-state observations, which may be delayed due to eventual consistency.
\end{itemize}

\subsection{Threat Model}
The security objective of \seb is to ensure that, within the deployment
assumptions above, production mutations occur only through broker
verification of authorized, fresh certificates. We consider an adversary
who can compromise the agent runtime, inject prompts, intercept network
packets, or manipulate the target state between the time of certificate
issuance and execution. Specifically, \seb defends against the following
threat vectors:
\begin{itemize}
    \item \textbf{Broker Bypass:} The agent attempts to directly invoke cloud APIs using standing credentials, bypassing both admission gates and the broker.
    \item \textbf{Stale Certificate Replay:} The adversary captures a previously executed certificate $\Omega$ and attempts to replay it to execute the mutation again.
    \item \textbf{Request-Certificate Mismatch:} The agent receives a certificate for a safe mutation (e.g., restarting a service) but submits a request for a destructive operation (e.g., terminating the database).
    \item \textbf{Time-of-Check to Time-of-Use (TOCTOU) Drift:} The state of the infrastructure changes between the time \sab issues the certificate and when \seb receives it, rendering the original contract dangerous.
    \item \textbf{Stale Epoch / Revocation Bypass:} The certificate is revoked or the system's global security policies are upgraded, but the agent tries to use an older certificate.
    \item \textbf{Overbroad Credential Abuse:} The agent attempts to leverage session credentials minted for a specific resource to mutate unrelated resources.
    \item \textbf{Broker Partition:} The agent attempts to execute mutations during a network partition that prevents the broker from contacting the Revocation Service.
\end{itemize}

%% file: sections/04-execution-model.tex
\section{Broker Execution Model}
\label{sec:execution-model}

The core of the \seb architecture is a formal verification and execution pipeline over certificates, execution requests, live target state, platform scopeability, signed DecisionRecords, and signed OutcomeRecords.

\subsection{Broker Execution Interface}
The Sovereign Execution Broker acts as a gatekeeper. It consumes an
SAB certificate $\Omega$, a concrete execution request
$req$, the live state of the target infrastructure $S_t$, and a platform
descriptor that captures the target API's native scoping mechanisms. The
interface is defined as:
\begin{equation}
    Execute(\Omega, req, S_t, Platform) \rightarrow D \quad \text{or} \quad O
\end{equation}
where $D$ is a signed \textsc{DecisionRecord} for verification rejection
or non-execution, and $O$ is a signed \textsc{OutcomeRecord} for a target
mutation attempt.

An admitted contract is represented as:
\begin{equation}
    C = (op, target, params, constraints, risk\_level)
\end{equation}
where $op$ is the intended API operation, $target$ is the resource or
endpoint to mutate, $params$ is the proposed payload, $constraints$
contains policy-level bounds on the mutation, and $risk\_level$ records
the assurance tier under which the action was admitted.

A SAB certificate is represented as:
\begin{equation}
\begin{split}
    \Omega ={} & (cid, C, E_{admit}, T_{valid}, P_{ver}, \rho_{rev}, \\
    & nonce, issuer, \sigma_{SAB})
\end{split}
\end{equation}
where $cid$ is the contract identifier, $E_{admit}$ is the evidence state
used during admission, $T_{valid}$ is the certificate validity interval,
$P_{ver}$ is the admitted policy version, $\rho_{rev}$ is the revocation
epoch at admission time, $nonce$ is a single-use freshness value,
$issuer$ identifies the \sab admission boundary, and $\sigma_{SAB}$ is the SAB
signature over the canonical certificate body.

An execution request $req$ is represented as a tuple:
\begin{equation}
    req = (cid, op, target, params, caller, t_{exec}, nonce)
\end{equation}
where:
\begin{itemize}
    \item $cid$: The unique identifier of the contract $C$ validated by \sab.
    \item $op$: The requested operation or API action.
    \item $target$: The target infrastructure resource or endpoint identifier.
    \item $params$: The exact parameters supplied to the API operation.
    \item $caller$: The identity of the agent wrapper requesting execution.
    \item $t_{exec}$: The timestamp of the execution request.
    \item $nonce$: The certificate nonce that must be unused in the broker ledger.
\end{itemize}

\begin{figure*}[t]
  \centering
  \begin{tikzpicture}[
    node distance=0.4cm and 0.6cm,
    block/.style={draw, rectangle, minimum height=0.7cm, minimum width=1.8cm, rounded corners, align=center, fill=blue!5, font=\scriptsize},
    arrow/.style={-Stealth, thick}
  ]
    \node (input) [block, fill=yellow!10] {Input\\$\Omega, req, S_t$};
    \node (sig) [block, right=of input] {$\Phi_{sig}$\\Signature};
    \node (match) [block, right=of sig] {$\Phi_{match}$\\Match};
    \node (time) [block, right=of match] {$\Phi_{time}$\\Validity};
    \node (policy) [block, right=of time] {$\Phi_{policy}$\\Policy Epoch};
    \node (rev) [block, below=of input, yshift=-0.5cm] {$\Phi_{rev}$\\Revocation};
    \node (drift) [block, right=of rev] {$\Phi_{drift}$\\Drift};
    \node (scope) [block, right=of drift] {$Scopeable$\\Platform};
    \node (mint) [block, right=of scope, fill=emerald!10] {Mint\\$ID_{exec}$};
    \node (api) [block, right=of mint] {API\\Mutation};
    \node (out) [block, right=of api, fill=orange!10] {Outcome\\Record $O$};
    \node (dec) [block, below=of scope, fill=red!10] {Decision\\Record $D$};

    \draw [arrow] (input) -- (sig);
    \draw [arrow] (sig) -- (match);
    \draw [arrow] (match) -- (time);
    \draw [arrow] (time) -- (policy);
    
    \draw [arrow] (policy.south) -- ++(0,-0.3) -| (rev.north);
    
    \draw [arrow] (rev) -- (drift);
    \draw [arrow] (drift) -- (scope);
    \draw [arrow] (scope) -- (mint);
    \draw [arrow] (scope) -- (dec);
    \draw [arrow] (mint) -- (api);
    \draw [arrow] (api) -- (out);
  \end{tikzpicture}
  \caption{Broker verification pipeline. The request $req$, certificate $\Omega$, and live state $S_t$ flow sequentially through cryptographic and state checks. Verification failures produce signed decision records; only if all checks pass is a short-lived execution identity minted and the API mutation attempted.}
  \label{fig:pipeline_flow}
\end{figure*}
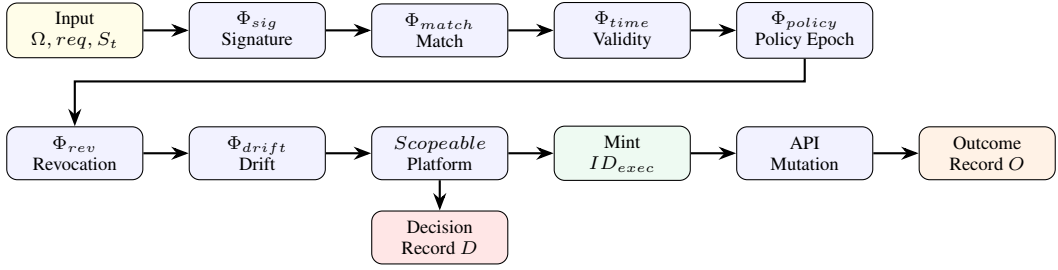

\subsection{Verification Pipeline}
When a request is submitted, the broker runs it through the pipeline shown in Figure~\ref{fig:pipeline_flow}. The stage predicates and their failure outcomes are summarized in Table~\ref{tab:checks}.

\input{sections/table-checks}

\subsection{Formal Certificate Verification Predicate}
Execution is admissible only if the verification predicate evaluates to
true. Let $V = VerifyExec(\Omega, req, S_t, Platform, L)$, where $L$ is
the broker ledger:
\begin{equation}
\begin{split}
    V ={} & \Phi_{sig} \land \Phi_{match} \land \Phi_{time} \\
    & \land \Phi_{policy} \land \Phi_{rev} \\
    & \land \Phi_{drift} \land \Phi_{replay} \\
    & \land \mathrm{Scopeable}(C, Platform)
\end{split}
\end{equation}
where:
\begin{itemize}
    \item $\Phi_{sig}$: Asserting the validity of \sab's cryptographic signature $\sigma_{SAB}$ on $\Omega$.
    \item $\Phi_{match}$: Confirming that $req.cid = \Omega.cid$, $req.nonce = \Omega.nonce$, and the requested operation, resource, and parameters match the contract $C$ embedded inside $\Omega$.
    \item $\Phi_{time}$: Confirming $t_{exec}$ lies within the certificate's validity window $T_{valid}$.
    \item $\Phi_{policy}$: Validating that the certificate policy version matches the active policy version: $\Phi_{policy} = (P_{ver} = P_{active})$.
    \item $\Phi_{rev}$: Confirming that the certificate has not been revoked ($\rho_{rev} = \rho_{active}$).
    \item $\Phi_{drift}$: Confirming that the live target state $S_t$ has not drifted beyond the admitted evidence state $E_{admit}$ described in the certificate.
    \item $\Phi_{replay}$: Confirming that $(cid, nonce)$ has not previously been reserved or executed in the broker ledger.
    \item $Scopeable(C, Platform)$: Confirming that the required action, resource, time, and parameter constraints are enforceable by the target platform or by a mandatory broker-controlled proxy/admission path.
\end{itemize}

Equivalently, the match predicate binds the request to the certificate
identifier and nonce before comparing the target contract:
\begin{equation}
\begin{split}
    \Phi_{match} ={} & (req.cid = \Omega.cid) \land \\
    & (req.nonce = \Omega.nonce) \land \\
    & Match(req.op, req.target, req.params, C).
\end{split}
\end{equation}
The policy predicate is:
\begin{equation}
    \Phi_{policy} = (P_{ver} = P_{active})
\end{equation}

The $Scopeable$ predicate is true only when every constraint in $C$ is
enforceable on the mutation path. Native IAM/RBAC conditions can make
action, resource, time, or namespace constraints scopeable. A mandatory
proxy or admission controller is another mechanism that makes
parameter-level constraints scopeable when native policy languages cannot
express them. If neither native policy nor a mandatory broker-controlled
interposition point can enforce a constraint, $Scopeable(C, Platform)$ is
false and the broker rejects before credential minting.

The replay predicate is:
\begin{equation}
    \Phi_{replay}(cid, nonce, L) = ((cid, nonce) \notin L_{used})
\end{equation}
The broker must atomically reserve $(cid, nonce)$ before minting an
execution identity or invoking the target API. A second request carrying
the same certificate nonce is rejected even if the first execution failed
after reservation.

\paragraph{Nonce Reservation, Idempotency, and Crash Recovery.}
Nonce reservation is conservative: once $(cid, nonce)$ is reserved, retry
is not allowed unless crash recovery can prove that no mutation was
attempted. A proof may come from a durable pre-invocation ledger state, an
idempotency token accepted by the target platform, or target-side evidence
showing that the mutation did not occur. If the broker crashes after the
point where target invocation may have happened, the certificate is
treated as consumed; the agent must re-admit the proposal through \sab
before another mutation attempt.

\subsection{DecisionRecord and OutcomeRecord Binding}
If verification fails before target invocation, \seb generates a
\textsc{DecisionRecord} $D$:
\begin{equation}
\begin{split}
    D ={} & \mathrm{DecisionRecord}(did, cid, nonce, decision, \\
    & failed\_check, reason, t_{decide}, \sigma_{broker})
\end{split}
\end{equation}
where $decision$ records whether the request was rejected, failed closed,
or required re-admission. Every verification failure produces a signed
DecisionRecord that binds the rejection reason to the certificate and
nonce.

If verification succeeds and the broker invokes the target mutation API,
\seb generates an \textsc{OutcomeRecord} $O$ representing the target attempt:
\begin{equation}
\begin{split}
    O ={} & \mathrm{OutcomeRecord}(oid, cid, nonce, status, \\
    & t_{start}, t_{end}, result, H(S_{post}), \\
    & error, \sigma_{broker})
\end{split}
\end{equation}
where $oid$ is the outcome ID, $status$ indicates target success or
target failure, $H(S_{post})$ is the hash of the target infrastructure's
post-state evidence when available, and $\sigma_{broker}$ is the broker's
digital signature binding the outcome to the original certificate and
nonce. OutcomeRecords are therefore evidence of attempted mutations, not
of verification failures.

\subsection{Failure Behavior}
If any check in the verification pipeline fails, \seb behaves according to defined policies:
\begin{itemize}
    \item \textbf{Reject:} Immediate rejection of execution request, returning a standard authorization error and writing a signed decision record.
    \item \textbf{Re-admit:} For transient failures like minor drift or policy mismatches, the broker writes a decision record and can request the agent to re-submit its proposal to \sab.
    \item \textbf{Fail Closed:} If the revocation status service is unreachable or if a signature check fails, \seb blocks the request and writes a fail-closed decision record.
    \item \textbf{Escalate:} Human break-glass handling is outside the autonomous certificate-bound path and is discussed in Section~\ref{sec:discussion}.
\end{itemize}

%% file: sections/table-checks.tex
\begin{table}[t]
\centering
\caption{Broker verification checks, inputs, and records emitted on failure.}
\label{tab:checks}
\small
\begin{tabular}{lll}
\toprule
\textbf{Check} & \textbf{Input} & \textbf{Failure Record} \\
\midrule
Signature $\Phi_{sig}$ & $\Omega, \sigma_{SAB}$ & $D$: reject \\
Contract match $\Phi_{match}$ & $req, \Omega, C$ & $D$: reject \\
Validity $\Phi_{time}$ & $T_{valid}$ & $D$: reject \\
Policy epoch $\Phi_{policy}$ & $P_{ver}, P_{active}$ & $D$: reject/re-admit \\
Revocation $\Phi_{rev}$ & $\rho_{rev}, \rho_{active}$ & $D$: fail closed \\
Drift $\Phi_{drift}$ & $E_{admit}, S_t$ & $D$: reject/re-admit \\
Replay $\Phi_{replay}$ & $cid, nonce, L$ & $D$: reject \\
Scopeability & $C, Platform$ & $D$: reject if false \\
\bottomrule
\end{tabular}
\end{table}

%% file: sections/05-scoped-identity.tex
\section{Scoped Execution Identity}
\label{sec:scoped-identity}

SEB shifts mutation authority from identity-centric authorization to certificate-bound scoped authorization.

\subsection{The Problem with Standing Credentials}
In typical cloud environments, agents or their runtime wrappers are assigned static IAM roles with generic permissions (e.g., \texttt{ec2:*} or \texttt{eks:*} privileges). Under this model, there is no enforcement mechanism preventing a wrapper or hijacked agent from bypassing the admission controller. An agent can issue a direct API request to mutate resources outside of any certified contract. To prevent bypass, we must enforce a strict constraint:
\begin{center}
    \emph{Agents and their wrappers must possess zero standing mutation credentials.}
\end{center}
Instead, the \seb is the only entity with the authority to assume powerful administrative roles.

\subsection{Bypass Prevention Deployment Pattern}
\label{sec:bypass-pattern}

SEB is an enforcement boundary only when it is installed as the mandatory
mutation path. The target platform must reject mutation calls from agent
identities, developer workstations, and ordinary service roles, and must
accept production mutation only from authorized mutation principals as
defined in Section~\ref{sec:system-threat}. Without this deployment
pattern, an agent with standing credentials can bypass both \sab and
\seb.

\paragraph{AWS.}
An AWS deployment uses a deny-by-default organization and account
configuration. Service Control Policies (SCPs) deny high-risk mutation
actions unless the principal is the broker role or a broker-issued
session carrying required session tags such as
\texttt{seb:cid}, \texttt{seb:nonce}, \texttt{seb:issuer}, and
\texttt{seb:risk}. Agent roles have permission boundaries that exclude
production mutation APIs and permit only proposal submission and
read-only evidence collection. The broker role alone may call
\texttt{sts:AssumeRole} into target execution roles. Those target roles
require broker-set session tags and external IDs, constrain session
duration to the certificate validity window where supported, and attach
inline STS session policies for action and resource restrictions. IAM
conditions and resource tags enforce native bounds such as action,
resource ARN, namespace/account, source VPC endpoint, and principal tag.
When AWS IAM cannot express a parameter-level constraint, such as a
specific ingress port or JSON payload field, the mutation must flow
through a broker-owned proxy adapter that validates the payload against
$C$ immediately before forwarding it to the AWS API and binding the
result to the ledger.

\paragraph{Kubernetes.}
A Kubernetes deployment separates agent service accounts from mutation
service accounts. Agent service accounts receive RBAC permissions for
proposal creation and observation but no verbs such as \texttt{create},
\texttt{update}, \texttt{patch}, or \texttt{delete} on protected
resources. The broker service account is the only account authorized to
perform those verbs in protected namespaces, and it uses short-lived
TokenRequest credentials or impersonation only after certificate
verification. A validating admission webhook rejects protected mutations
unless the request carries broker-attested context, such as
\texttt{seb.openkedge.io/cid} and \texttt{seb.openkedge.io/nonce}
annotations or equivalent admission-review metadata. A mutating webhook
may stamp broker-approved metadata onto objects before the validating
webhook checks it. Policy engines such as Gatekeeper, Kyverno, or
ValidatingAdmissionPolicy can enforce invariants that RBAC cannot
express, while the broker adapter validates payload fields that remain
outside native policy expressiveness. The security theorem below depends
on these RBAC and admission controls rejecting direct non-broker mutation
attempts.

\subsection{SEB Identity Model}
Rather than executing actions using its own master credentials, the \seb acts as a token broker. Upon verifying a certificate $\Omega$, the broker programmatically requests the Identity Provider (IdP) to mint a temporary, down-scoped execution credential $ID_{exec}$:
\begin{equation}
    ID_{exec} = MintIdentity(C, \Omega, target, scope, T_{valid}, \rho_{rev})
\end{equation}
Native IAM or RBAC systems enforce the dimensions they can express,
typically action, resource, principal, session duration, audience, and
namespace. Parameter-level constraints and certificate binding are not
uniformly enforceable by cloud IAM. When the target API cannot enforce a
dimension directly, \seb preserves the binding through broker-side proxy
validation, admission control, required session tags, and ledger binding.

\input{sections/table-dimensions}

\subsection{Scope Dimensions}
The down-scoped identity restricts execution along several dimensions.
These dimensions are detailed in Table~\ref{tab:dimensions}. For
instance, in the firewall example, AWS session policy can restrict the
session to \texttt{ec2:AuthorizeSecurityGroupIngress} on the target
security-group ARN and to a short duration, while session tags bind the
credential to the certificate identifier and nonce. If IAM cannot express
the exact port or CIDR constraint for the relevant API, the broker-owned
proxy adapter validates \texttt{port=443} and the approved CIDR against
$C$ before forwarding the call; the ledger then records the certificate,
nonce, request payload, and outcome. Thus native authorization enforces
what the platform supports, and \seb supplies the reference-monitor logic
for the remaining certificate-bound constraints.

\begin{figure}[t]
  \centering
  \begin{tikzpicture}[
    node distance=0.6cm,
    block/.style={draw, rectangle, minimum height=0.6cm, minimum width=3.2cm, rounded corners, align=center, fill=blue!5, font=\footnotesize},
    subblock/.style={draw, rectangle, minimum height=0.5cm, minimum width=2.8cm, align=left, fill=gray!10, font=\scriptsize},
    arrow/.style={-Stealth, thick}
  ]
    \node (omega) [block, fill=emerald!10] {Assurance Certificate $\Omega$};
    \node (c) [subblock, below=of omega, yshift=0.2cm] {$\bullet$ Contract $C$};
    \node (res) [subblock, below=of c, yshift=0.3cm] {$\bullet$ Target Resource};
    \node (op) [subblock, below=of res, yshift=0.3cm] {$\bullet$ Action \& Parameters};
    \node (time) [subblock, below=of op, yshift=0.3cm] {$\bullet$ Validity Window $T_{valid}$};
    \node (epoch) [subblock, below=of time, yshift=0.3cm] {$\bullet$ Revocation Epoch $\rho_{rev}$};
    
    \node (identity) [block, below=of epoch, yshift=-0.4cm, fill=orange!10] {Short-Lived Execution Identity\\$ID_{exec}$};

    \draw [arrow] (omega) -- (c);
    \node (box) [draw, dashed, fit=(c) (epoch), inner sep=0.12cm, label=right:\scriptsize Scope Bindings] {};
    
    \draw [arrow] (box.south) -- (identity);
  \end{tikzpicture}
  \caption{Scoped identity binding. The broker extracts the safety parameters, target resources, and validity windows from $\Omega$ and maps them into a short-lived execution identity $ID_{exec}$ plus any required broker-side enforcement bindings.}
  \label{fig:scoped_identity}
\end{figure}
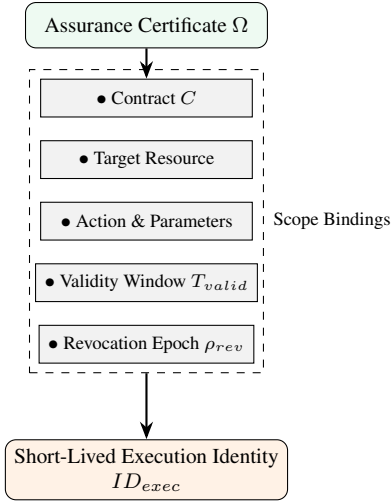

\subsection{Security Properties}
The scoped execution identity model provides several security properties under the deployment assumptions in Section~\ref{sec:system-threat}:
\begin{itemize}
    \item \textbf{No Standing Credentials:} Agents never hold credentials that can be abused out-of-context.
    \item \textbf{Least Privilege:} Tokens are restricted to the minimum required API permissions for the requested task.
    \item \textbf{Constrained Reuse:} Tokens are bound to a short expiration window (60 seconds for Kubernetes projected tokens, and the minimum supported 900 seconds for AWS STS). While nonce reservation prevents repeated broker-mediated minting, exposure of an already minted 900-second AWS STS credential is bounded by session policies, session tags, the proxy path, and target IAM conditions, preventing out-of-context reuse.
    \item \textbf{Revocation-Before-Execution:} If the certificate is revoked before execution, the broker will refuse to mint the token, preventing the action from occurring even if it was previously approved.
    \item \textbf{Replay Resistance:} The broker reserves the certificate nonce before mutation and rejects subsequent requests carrying the same $(cid, nonce)$ pair.
    \item \textbf{Accountability:} Every broker decision and mutation attempt is logged, providing cryptographic evidence linking the executor's identity to the original SAB certificate, decision record, and outcome record where a target invocation occurred.
\end{itemize}

%% file: sections/table-dimensions.tex
\begin{table}[t]
\centering
\caption{Scoped identity dimensions and their enforcement mechanisms.}
\label{tab:dimensions}
\small
\begin{tabular}{lp{4.9cm}}
\toprule
\textbf{Dimension} & \textbf{Enforcement} \\
\midrule
Action & Native IAM/RBAC operation where supported (e.g., AWS security-group ingress) \\
Resource & Native resource ARN, namespace, object name, or label selector \\
Parameters & Native conditions, or mandatory broker proxy/admission validation against $C$ \\
Time & STS duration, projected-token expiration, and broker validity check over $T_{valid}$ \\
Certificate & Session tags, admission metadata, and ledger binding to $\Omega$ and $nonce$ \\
Policy & Broker validation that $P_{ver}=P_{active}$ \\
Revocation & Broker validation that $\rho_{rev}=\rho_{active}$ before minting \\
\bottomrule
\end{tabular}
\end{table}

%% file: sections/06-drift-revocation-toctou.tex
\section{Drift, Revocation, and TOCTOU Protection}
\label{sec:drift-revocation}

Distributed control planes introduce a vulnerable interval between action admission and runtime execution. SEB closes that interval with live-state drift checks and epoch-based revocation.

\subsection{The TOCTOU Problem in Agentic Systems}
When \sab evaluates an agent proposal at time $t_1$, it does so based on the admitted evidence state $E_{admit}=E_{t_1}$ gathered at that moment. SAB certifies that the proposed contract $C$ is safe under $E_{admit}$ and signs the certificate $\Omega$. However, the certificate may not be presented to the \seb for execution until a later time $t_2$ (where $t_2 > t_1$). 

During this interval $t_2 - t_1$, the target infrastructure can change. For example, another process may modify security group rules, rotate access keys, or decommission target virtual machines. If the broker executes the certificate blindly without checking for state changes, a certified action that was safe at $t_1$ could become unsafe or policy-violating at $t_2$.

\subsection{Drift Predicate}
To prevent execution on stale state, \seb implements a live drift check. Before executing any mutation, the broker queries the target infrastructure's current state $S_{t_2}$ and compares it to the evidence state $E_{admit}$ captured inside $\Omega$. We define the drift predicate $\Phi_{drift}$ as:
\begin{equation}
\begin{split}
    \Phi_{drift}(E_{admit}, S_{t_2}) ={} & Distance(Project(E_{admit}), \\
    & Project(S_{t_2})) \le \epsilon_C
\end{split}
\end{equation}
where:
\begin{itemize}
    \item $Project(\cdot)$: A projection function extracting only the state variables relevant to contract $C$.
    \item $Distance(\cdot)$: A distance or divergence metric measuring the discrepancy between the admitted state and the live state.
    \item $\epsilon_C$: The policy-defined drift tolerance limit for contract $C$.
\end{itemize}
If the distance exceeds $\epsilon_C$, the broker rejects the certificate, rendering the action invalid.

\subsection{Epoch-Based Revocation}
In addition to target state changes, corporate policy or operational requirements may change after a certificate is signed. Rather than keeping track of long, distributed revocation lists (CRL), we employ a global revocation epoch counter $\rho_{active}$. 

\begin{figure}[t]
  \centering
  \begin{tikzpicture}[
    node distance=1.0cm,
    timepoint/.style={circle, draw, fill=blue!10, inner sep=2pt, font=\scriptsize},
    labelnode/.style={align=center, font=\scriptsize},
    arrow/.style={-Stealth, thick}
  ]
    \draw [thick, -Stealth] (0,0) -- (7.8,0) node [right] {$t$};
    
    \node (t1) at (0.8,0) [timepoint] {};
    \node (t2) at (2.8,0) [timepoint, fill=red!20] {};
    \node (t3) at (4.8,0) [timepoint] {};
    \node (t4) at (6.8,0) [timepoint, fill=red!40] {};

    \node at (0.8, 0.4) [labelnode] {$t_1$};
    \draw [gray, dashed, thin] (t1.south) -- (0.8, -0.4);
    \node at (0.8, -0.4) [labelnode, anchor=north] {SAB\\admits $C$,\\signs $\Omega$ ($\rho_{rev}=k$)};

    \node at (2.8, 0.4) [labelnode] {$t_2$};
    \draw [gray, dashed, thin] (t2.south) -- (2.8, -1.2);
    \node at (2.8, -1.2) [labelnode, anchor=north] {Revocation\\epoch advances\\to $k+1$};

    \node at (4.8, 0.4) [labelnode] {$t_3$};
    \draw [gray, dashed, thin] (t3.south) -- (4.8, -0.4);
    \node at (4.8, -0.4) [labelnode, anchor=north] {Agent requests\\execution of $\Omega$};

    \node at (6.8, 0.4) [labelnode] {$t_4$};
    \draw [gray, dashed, thin] (t4.south) -- (6.8, -1.2);
    \node at (6.8, -1.2) [labelnode, anchor=north] {Broker rejects\\$\Omega$ (stale $\rho_{rev}$)};
  \end{tikzpicture}
  \caption{Revocation-before-execution timeline. Between admission ($t_1$) and execution request ($t_3$), the active revocation epoch advances ($t_2$), causing the broker to reject the request ($t_4$).}
  \label{fig:revocation_timeline}
\end{figure}
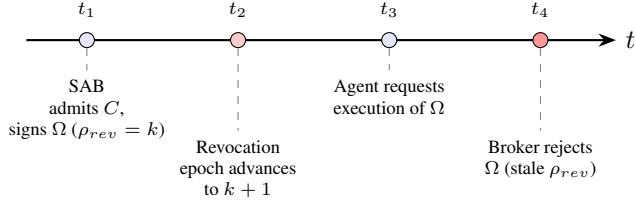

Each certificate $\Omega$ contains the revocation epoch $\rho_{rev}$ at which it was signed. At the moment of execution, the broker verifies:
\begin{equation}
    \rho_{active} = \rho_{rev} \Rightarrow \text{executable}
\end{equation}
If the system has advanced the global epoch (e.g., due to an active security incident or policy change) such that:
\begin{equation}
    \rho_{active} > \rho_{rev} \Rightarrow \text{Reject}
\end{equation}
the certificate is rejected immediately. Figure~\ref{fig:revocation_timeline} shows the timeline of this process.

\subsection{Fail-Closed Semantics}
Under high-assurance levels (specifically L3/L4 within the \openkedge framework), the broker enforces a strict fail-closed policy. In the event of system partitions, network latency, or service failures, \seb defaults to rejecting the request. Specifically:
\begin{itemize}
    \item \textbf{Unreachable Revocation Service:} If \seb cannot query the active global epoch $\rho_{active}$, it assumes $\rho_{active} > \rho_{rev}$ and rejects.
    \item \textbf{Drift Evaluation Failure:} If target APIs fail to return live state metrics or evidence is incomplete, \seb rejects the transaction.
    \item \textbf{Stale Policy Epoch:} If the current policy version exceeds the policy epoch indicated in $\Omega$, the request is rejected and forced to undergo re-admission.
\end{itemize}
This fail-closed design prevents an attacker from isolating the broker to exploit older, compromised certificates.

%% file: sections/07-implementation.tex
\section{System Implementation}
\label{sec:implementation}

We have implemented a concrete research prototype of the Sovereign Execution Broker (\seb) in Go, consisting of approximately 4,200 lines of code. The codebase is structured into modular packages with clean boundaries to separate cryptographic verification, state monitoring, ledger writing, and platform-specific credential minting. A compact summary of the system implementation artifacts is presented in Table~\ref{tab:artifact_summary}.

\begin{table}[t]
\centering
\caption{Compact summary of the \seb research prototype artifacts.}
\label{tab:artifact_summary}
\scriptsize
\begin{tabular}{lp{4.8cm}}
\toprule
\textbf{System Aspect} & \textbf{Value / Implementation Choice} \\
\midrule
Language & Go v1.21 \\
Lines of Code (LoC) & ~4,200 LoC (excluding tests) \\
Broker Replicas & 3 replicas (high-availability mode) \\
Database Backend & PostgreSQL v15 (Amazon RDS \texttt{db.r6g.xlarge}) \\
Scoped Identity Adapters & AWS STS, Kubernetes TokenRequest \\
Enforcement Components & AWS Mutation Proxy, Kubernetes Validating Webhook \\
Cryptographic Library & Go standard library \texttt{crypto/ed25519} \\
Revocation Caching & Local \texttt{sync.Map}, 5-sec poll interval, 5-sec TTL \\
Deployment Environment & AWS EKS v1.28 (worker instances: \texttt{m6i.xlarge}) \\
\bottomrule
\end{tabular}
\end{table}

\begin{figure*}[t]
  \centering
  \begin{tikzpicture}[
    node distance=0.8cm and 1.2cm,
    block/.style={draw, rectangle, minimum height=0.8cm, minimum width=2.4cm, rounded corners, align=center, fill=blue!5, font=\scriptsize},
    database/.style={draw, rectangle, rounded corners, minimum height=0.8cm, minimum width=1.5cm, align=center, fill=gray!10, font=\scriptsize},
    arrow/.style={-Stealth, thick}
  ]
    \node (client) [block] {Agent / Wrapper\\gRPC Request};
    \node (broker) [block, right=of client, fill=emerald!10] {Go Broker Service\\(Ed25519 Verifier)};
    \node (cache) [block, above=of broker] {Revocation Epoch\\Cache};
    \node (sts) [block, below=of broker, fill=orange!10] {Cloud STS Adapter\\(Session Policies)};
    \node (ledger) [database, right=of broker] {Ledger\\(PostgreSQL)};
    \node (api) [block, below=of ledger] {Cloud / Kubernetes\\APIs (Mutation)};

    \draw [arrow] (client) -- (broker);
    \draw [arrow] (broker) -- (cache);
    \draw [arrow] (broker) -- (sts);
    \draw [arrow] (sts) -- (api);
    \draw [arrow] (broker) -- (ledger);
    \draw [arrow] (api.north) -- node[right] {\scriptsize Outcome Binding} (ledger.south);
  \end{tikzpicture}
  \caption{System deployment architecture of the Sovereign Execution Broker. The Go-based service verifies certificates, interacts with active local caches, calls cloud STS or Kubernetes APIs for scoped session credentials, invokes mutations through broker-controlled adapters/proxies, and logs signed decision and outcome records to PostgreSQL.}
  \label{fig:prototype_architecture}
\end{figure*}
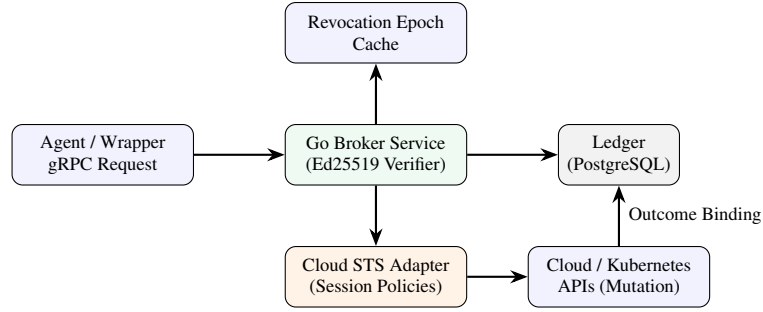

\subsection{Codebase Structure and Boundaries}
The broker's codebase is divided into the following primary components:
\begin{itemize}
    \item \textbf{\texttt{cmd/seb-broker}:} The entry-point daemon that hosts the gRPC and HTTP/JSON endpoints.
    \item \textbf{\texttt{pkg/verifier}:} The verification pipeline that parses certificates, verifies signatures, checks validity intervals, and matches requests to contracts.
    \item \textbf{\texttt{pkg/revocation}:} The polling client that interacts with the Revocation Epoch Service and maintains the local thread-safe epoch cache.
    \item \textbf{\texttt{pkg/ledger}:} The database client managing PostgreSQL storage for nonces, decision records, and outcome records.
    \item \textbf{\texttt{pkg/adapter}:} Platform-specific adapters including the AWS Security Token Service (STS) and Kubernetes TokenRequest adapters.
    \item \textbf{\texttt{pkg/proxy}:} The reverse proxy for parameter-level API validation.
\end{itemize}

\subsection{Certificate Parser and Verifier}
The certificate verifier consumes SAB certificates ($\Omega$) and validates their authenticity. We use Go's standard \texttt{crypto/ed25519} library for cryptographic operations. To prevent signature-malleability exploits, certificates are parsed and checked against a strict JSON schema, then marshaled using a deterministic Canonical JSON serialization (sorting keys alphabetically, removing optional whitespace, and escaping strings per RFC 8785) before signature verification. The verifier loads SAB's public verification key from a secure local trust store, extracts the contract $C$, and validates the signature $\sigma_{SAB}$ over the canonical certificate body.

\subsection{Nonce Reservation and Replay Protection}
To prevent replay attacks, \seb implements an atomic nonce reservation mechanism using a PostgreSQL database. When an execution request is received, the broker initiates a database transaction and attempts to insert the \texttt{(cid, nonce)} pair into a \texttt{reserved\_nonces} table:
\begin{verbatim}
CREATE TABLE reserved_nonces (
    cid UUID NOT NULL,
    nonce VARCHAR(64) NOT NULL,
    reserved_at TIMESTAMP NOT NULL,
    PRIMARY KEY (cid, nonce)
);
\end{verbatim}
Because of the primary key constraint, duplicate requests fail the insertion and are immediately rejected. This prevents double-spend attacks even under highly concurrent execution requests.

\subsection{Decision and Outcome Ledger}
Every verification decision and execution outcome is logged to PostgreSQL. The schemas are defined as:
\begin{verbatim}
CREATE TABLE decision_records (
    did UUID PRIMARY KEY,
    cid UUID NOT NULL,
    nonce VARCHAR(64) NOT NULL,
    decision VARCHAR(32) NOT NULL,
    failed_check VARCHAR(32),
    reason TEXT,
    t_decide TIMESTAMP NOT NULL,
    signature BYTEA NOT NULL
);

CREATE TABLE outcome_records (
    oid UUID PRIMARY KEY,
    cid UUID NOT NULL,
    nonce VARCHAR(64) NOT NULL,
    status VARCHAR(32) NOT NULL,
    t_start TIMESTAMP NOT NULL,
    t_end TIMESTAMP NOT NULL,
    result TEXT,
    post_state_hash VARCHAR(64),
    error TEXT,
    signature BYTEA NOT NULL
);
\end{verbatim}
Every record is signed by the broker using its private Ed25519 key, ensuring non-repudiation and providing an audit trail for security teams.

\subsection{Policy and Revocation Cache}
The broker maintains a thread-safe local cache of the global policy version $P_{active}$ and revocation epoch $\rho_{active}$ using a \texttt{sync.Map}. A background worker polls the centralized gRPC Revocation Service over a TLS connection every 5 seconds. To enforce fail-closed behavior, the local cache has a TTL of 5 seconds. If a network partition prevents the broker from contacting the revocation service, the cache expires after 5 seconds, and all subsequent certificate validations fail closed until connectivity is restored.

\subsection{Scoped Identity Adapters}
The broker implements two credential-minting adapters:
\begin{itemize}
    \item \textbf{AWS STS Adapter:} Evaluates the contract $C$ and calls the AWS STS \texttt{AssumeRole} API, requesting the minimum supported session duration of 900 seconds. While nonce reservation prevents repeated broker-mediated minting, exposure of an already minted 900-second AWS STS credential is bounded by session policies, session tags, proxy path, and target IAM conditions, preventing out-of-context reuse.
    \item \textbf{Kubernetes TokenRequest Adapter:} Invokes the Kubernetes core token projection API (\texttt{/api\slash v1\slash namespaces\slash \{ns\}\slash serviceaccounts\slash \{sa\}\slash token}) via the Go \texttt{client-go} SDK. It requests a short-lived token (expiration window of 60 seconds) bound to the target namespace, service account, and audience.
\end{itemize}

\subsection{Cloud Proxy and Admission Webhooks}
To enforce constraints that native IAM and RBAC cannot express (such as parameter-level validation like restricting port ranges or container images), \seb implements inline enforcement components:
\begin{itemize}
    \item \textbf{AWS Mutation Proxy:} A secure HTTP/HTTPS reverse proxy that intercepts mutation requests. The agent wrapper directs its API requests to this proxy, which uses the dynamically minted scoped STS credentials. The proxy parses the request payload, validates parameter constraints against the contract $C$ (e.g., verifying that the CIDR block and ports match the certified contract), and forwards the request to the cloud endpoint.
    \item \textbf{Kubernetes Validating Webhook:} A validating admission webhook service (implemented using Go's \texttt{sigs.k8s.io/controller-runtime}) that intercepts Kubernetes resource modifications. The webhook rejects any mutation that does not carry the \texttt{seb.openkedge.io/cid} and \texttt{seb.openkedge.io/nonce} annotations matching a successfully reserved nonce and valid certificate in the broker's ledger.
\end{itemize}

\subsection{Trusted Computing Assumptions and Failure Handling}
\textbf{Trusted Computing Base (TCB):} We assume the broker runs in a secure, isolated execution environment (e.g., AWS Nitro Enclaves or a dedicated Kubernetes private network segment). We trust the integrity of the broker's private signing key, the SAB public key store, the PostgreSQL database, and the target platform's root IAM/RBAC mechanisms.

\textbf{Failure and Crash Recovery:} If the broker crashes after nonce reservation but before invoking the target API, the execution request fails closed, and the nonce remains consumed. The agent must re-request admission from SAB to obtain a fresh certificate. To prevent double mutation, the broker passes an \emph{idempotency token} (computed as $H(cid \mathbin{\Vert} nonce)$) to the target platform (e.g., via the AWS \texttt{ClientRequestToken} parameter or Kubernetes client uid). Upon broker recovery, if the status of a reserved nonce is uncertain, the broker queries the target API using the idempotency token to determine if the mutation occurred, logging the result before releasing the transaction lock. Where target APIs do not expose reliable idempotency or lookup mechanisms, \seb conservatively treats the certificate as consumed and requires re-admission through \sab.

\subsection{Artifact Availability}
The \seb research prototype, including the Go gRPC and HTTP broker service, AWS and Kubernetes adapters, validating webhook manifests, benchmark harness scripts, and detailed reproducibility instructions, will be released publicly as open-source software as part of the OpenKedge artifact repository.

%% file: sections/08-evaluation.tex
\section{System Evaluation}
\label{sec:evaluation}

We evaluated our Go-based \seb research prototype to assess its latency overhead, throughput scalability, revocation propagation delay, and enforcement capability under injected security threats and system faults. Our evaluation answers the following research questions:
\begin{itemize}
    \item \textbf{RQ1 (Latency Overhead):} What latency overhead does \seb add to infrastructure mutations?
    \item \textbf{RQ2 (Revocation Propagation):} How quickly does a global revocation epoch advancement prevent pending executions?
    \item \textbf{RQ3 (Credential Minting Cost):} What is the computational and network cost of dynamic scoped credential issuance?
    \item \textbf{RQ4 (Drift Detection):} How does the drift check perform under live state queries, and what is its overhead?
    \item \textbf{RQ5 (Comparative Advantage):} How does broker enforcement compare to direct IAM, static policy checks, or audit-only configurations?
    \item \textbf{RQ6 (Fault Containment):} How does \seb behave under network partitions, broker crashes, or malformed certificate attacks?
\end{itemize}

\input{sections/table-metrics}

\subsection{Experimental Setup}
Our experimental setup consists of a Kubernetes cluster (Amazon EKS v1.28) deployed in the AWS \texttt{us-west-2} region.
\begin{itemize}
    \item \textbf{Broker Nodes:} The Go-based broker service runs as a high-availability deployment with 3 replicas. Each replica is hosted on an AWS EC2 \texttt{m6i.xlarge} instance (4 vCPUs, 16 GiB RAM) running Linux.
    \item \textbf{Database Backend:} We use Amazon RDS PostgreSQL v15 (\texttt{db.r6g.xlarge}) as our ledger and nonce storage backend.
    \item \textbf{Network Topology:} All broker nodes, the database instance, and the Revocation Service reside within a private VPC. The average network latency between broker instances and the RDS PostgreSQL instance is 1.1 ms.
    \item \textbf{Certificates:} SAB certificates are generated using Ed25519 signatures, ranging in size from 512 bytes (minimal IAM resource policy) to 1.5 KB (complex policies containing multiple parameter-level constraints).
    \item \textbf{Revocation Cache:} The local cache polls the gRPC Revocation Service every 5 seconds, with a cache TTL of 5 seconds.
    \item \textbf{Credential Duration:} Projected service account tokens are requested with a 60-second duration, and AWS STS credentials are minted with the minimum allowed session duration of 900 seconds.
\end{itemize}

\paragraph{Methodology and Simulation Boundary.} All latency and throughput measurements for the baseline configurations and Full SEB were collected using live cloud and database APIs in AWS. Specifically, our broker-side operations interacted directly with active AWS STS AssumeRole endpoints, the live Kubernetes core TokenRequest API, and our RDS PostgreSQL database instance. The live drift checks performed real queries to active AWS EC2 Security Groups and EKS cluster APIs. Conversely, for the security and fault-injection experiments (reported in Section 8.5), network partitions of the revocation service and database crash events were simulated within our controlled test-harness environment to prevent permanent disruption to our active AWS cloud infrastructure.

\paragraph{Statistical Methodology.} To ensure statistical rigor, we adopted a strict measurement policy. For each configuration and workload, we ran 100 warm-up execution cycles to prime Go runtime caches, connection pools, TLS sessions, and database/client-side caches, after which we executed 5,000 official trials. Tables~\ref{tab:microbenchmarks}--\ref{tab:breakdown} report latency measurements collected at a fixed concurrency level of a single-request steady-state (concurrency = 1) over 5,000 trials. To evaluate scalability, we conducted a separate experiment scaling request concurrency from 1 to 100 threads, with the throughput results reported in Figure~\ref{fig:throughput}. The latency values are summarized using p50, p95, and p99 percentiles rather than simple arithmetic averages, capturing tail latency anomalies. The AWS SDK for Go was configured such that transient network retries were enabled up to three times, while retry-on-throttling was disabled so STS rate limits appeared directly in throughput measurements.

\subsection{Verification Microbenchmarks (Exp 1)}
We first measured the execution latency of each check in the verification pipeline. Table~\ref{tab:microbenchmarks} reports the p50 and p99 latencies over 5,000 trials at a fixed single-request steady-state concurrency (concurrency = 1).

\begin{table}[t]
\centering
\caption{Broker verification microbenchmarks (5,000 trials).}
\label{tab:microbenchmarks}
\scriptsize
\begin{tabular}{p{3.3cm}rr}
\toprule
\textbf{Pipeline Component} & \textbf{\shortstack{p50 Latency\\(ms)}} & \textbf{\shortstack{p99 Latency\\(ms)}} \\
\midrule
Ed25519 Signature Verification ($\Phi_{sig}$) & 0.08 & 0.11 \\
Certificate Parsing \& Canonicalization & 0.12 & 0.18 \\
Contract Matching ($\Phi_{match}$) & 0.05 & 0.08 \\
Epoch Check ($\Phi_{policy}, \Phi_{rev}$) (Cache Hit) & 0.03 & 0.05 \\
Epoch Check ($\Phi_{policy}, \Phi_{rev}$) (Cache Miss) & 2.40 & 3.82 \\
PostgreSQL Nonce Reservation ($\Phi_{replay}$) & 3.20 & 5.12 \\
PostgreSQL Ledger Write (Async Append) & 3.50 & 5.60 \\
AWS STS Credential Minting (\texttt{AssumeRole}) & 46.70 & 158.40 \\
Kubernetes TokenRequest API & 12.10 & 18.30 \\
\bottomrule
\end{tabular}
\end{table}

The cryptographic validation ($\Phi_{sig}$, parsing, and contract matching) contributes less than 0.3 ms of total latency. The primary latency bottlenecks are the network round-trips to external services: the database write for nonce reservation (~3.2 ms) and, most dominantly, AWS STS credential minting (~105.2 ms).

\subsection{End-to-End Latency and Baselines (Exp 2)}
We compared the end-to-end performance of \seb against several baselines across two workloads:
\begin{itemize}
    \item \textbf{K8s Patch:} Updating pod replicas or container configurations in the EKS cluster.
    \item \textbf{AWS Security Group Update:} Intercepting and executing an ingress firewall rule authorization.
\end{itemize}
We report the baseline latency measurements in Table~\ref{tab:baselines_k8s} (Kubernetes workload) and Table~\ref{tab:baselines_aws} (AWS workload), separating the broker-side overhead from the estimated client-observed end-to-end latency. Note that SAB admission latency is excluded from all configurations because the evaluation isolates mutation-time enforcement overhead.

\begin{table}[h]
\centering
\caption{Kubernetes workload latency comparison (p50, p95, p99) showing broker-only overhead and estimated client end-to-end latency.}
\label{tab:baselines_k8s}
\scriptsize
\begin{tabular}{p{3.0cm}cc}
\toprule
\textbf{Configuration} & \textbf{\shortstack{Broker-Only\\(ms)}} & \textbf{\shortstack{Estimated\\Client E2E\\(ms)}} \\
\midrule
Direct RBAC & 0.00 / 0.00 / 0.00 & 12.50 / 15.20 / 19.80 \\
Static Policy Proxy (Gatekeeper) & 8.50 / 11.20 / 12.30 & 21.00 / 26.40 / 32.10 \\
Audit-Only Logging & 0.00 / 0.00 / 0.00 & 12.70 / 15.40 / 20.10 \\
SAB without Broker Enforcement & 0.00 / 0.00 / 0.00 & 12.60 / 15.30 / 19.90 \\
\midrule
\textbf{Full SEB (Ours)} & \textbf{28.20 / 36.50 / 47.10} & \textbf{40.70 / 51.70 / 66.90} \\
-- SEB without drift checks & 16.10 / 21.30 / 27.80 & 28.60 / 36.50 / 47.60 \\
-- SEB without revocation checks & 28.10 / 36.40 / 47.00 & 40.60 / 51.60 / 66.80 \\
-- SEB without nonce reservation & 25.00 / 32.20 / 41.50 & 37.50 / 47.40 / 61.30 \\
\bottomrule
\end{tabular}
\end{table}

\begin{table}[h]
\centering
\caption{AWS workload latency comparison (p50, p95, p99) showing broker-only overhead and estimated client end-to-end latency.}
\label{tab:baselines_aws}
\scriptsize
\begin{tabular}{p{2.4cm}cc}
\toprule
\textbf{Configuration} & \textbf{\shortstack{Broker-Only\\(ms)}} & \textbf{\shortstack{Estimated\\Client E2E\\(ms)}} \\
\midrule
Direct IAM & 0.00 / 0.00 / 0.00 & 85.00 / 98.40 / 112.50 \\
Static Policy Proxy (OPA) & 8.20 / 10.10 / 12.10 & 93.20 / 108.50 / 124.60 \\
Audit-Only Logging & 0.00 / 0.00 / 0.00 & 86.10 / 99.70 / 114.20 \\
SAB without Broker Enforcement & 0.00 / 0.00 / 0.00 & 85.40 / 98.90 / 113.10 \\
\midrule
\textbf{Full SEB (Ours)} & \textbf{136.90 / 180.40 / 284.10} & \textbf{221.90 / 278.80 / 396.60} \\
-- SEB without drift checks & 51.60 / 81.90 / 166.10 & 136.60 / 180.30 / 278.60 \\
-- SEB without revocation checks & 136.80 / 180.20 / 283.90 & 221.80 / 278.60 / 396.40 \\
-- SEB without nonce reservation & 133.70 / 175.90 / 279.00 & 218.70 / 274.30 / 391.50 \\
\bottomrule
\end{tabular}
\end{table}

The measurements are separated into two distinct metrics:
\begin{itemize}
    \item \textbf{Broker-Only Overhead:} The total latency added solely by the broker service, comprising certificate signature verification, database nonce reservation, live drift check query, scoped credential minting (STS or TokenRequest), proxy payload validation, and database ledger write.
    \item \textbf{Estimated Client End-to-End Latency:} The total operation time observed by the client application, which is estimated by adding the isolated broker-only overhead to the baseline direct target API execution latency (12.50 ms for Kubernetes, 85.00 ms for AWS).
\end{itemize}
This separation prevents the subtraction of target API baselines from the broker's isolated overhead.

While Direct IAM and Audit-Only configurations show the lowest latency (as they perform no runtime checks or token minting), they fail to protect against agent bypass or credential theft. SEB adds 28.2 ms broker-side overhead to Kubernetes mutations and 136.9 ms broker-side overhead to AWS security-group mutations; the resulting estimated client-observed p50 latencies are approximately 40.7 ms and 221.9 ms when combined with direct target API latency.

The ablation experiments reveal the source of this overhead. In AWS, removing the live drift check (\emph{SEB without drift}) reduces the broker-only overhead by 85.3 ms because it eliminates the need to query the AWS API (\texttt{DescribeSecurityGroups}) to verify the subnet state. Removing the database nonce reservation (\emph{SEB without nonce reservation}) saves 3.2 ms of database write time.

\subsection{Latency Breakdown Analysis}
To decompose the performance overhead of the broker, we analyze the latency contributions of individual sub-steps in Table~\ref{tab:breakdown}.

\begin{table*}[t]
\centering
\caption{Latency breakdown of Full SEB execution overhead.}
\label{tab:breakdown}
\scriptsize
\begin{tabular}{lrrrr}
\toprule
 & \multicolumn{2}{c}{\textbf{K8s Workload}} & \multicolumn{2}{c}{\textbf{AWS Workload}} \\
\cmidrule(lr){2-3} \cmidrule(lr){4-5}
\textbf{Pipeline Component} & \textbf{Latency (ms)} & \textbf{Share (\%)} & \textbf{Latency (ms)} & \textbf{Share (\%)} \\
\midrule
Verification ($\Phi_{sig}$, parsing, matching) & 0.25 & 0.9 & 0.28 & 0.2 \\
Nonce Reservation ($\Phi_{replay}$) & 3.20 & 11.3 & 3.20 & 2.3 \\
Drift Query ($\Phi_{drift}$) & 12.10 & 42.9 & 85.30 & 62.3 \\
Credential Minting ($ID_{exec}$) & 12.10 & 42.9 & 46.70 & 34.1 \\
Ledger Write (Post-execution logging) & 0.55 & 2.0 & 1.42 & 1.1 \\
\midrule
\textbf{Total Overhead} & \textbf{28.20} & \textbf{100.0} & \textbf{136.90} & \textbf{100.0} \\
\bottomrule
\end{tabular}
\end{table*}

This breakdown demonstrates that target state drift checking (42.9\% for Kubernetes, 62.3\% for AWS) and scoped credential minting (42.9\% for Kubernetes, 34.1\% for AWS) constitute the vast majority of the overhead. In contrast, local validation logic represents a negligible portion of the overhead.

\subsection{Security and Failure-Mode Validation (Exp 3 \& 5)}
We evaluated the broker's resilience through a controlled injected test suite of security threat vectors and system faults, executing 1,000 test cases for each scenario. Table~\ref{tab:security} reports the results.

\begin{table*}[t]
\centering
\caption{Security and failure-mode controlled injected test-suite results (1,000 cases each).}
\label{tab:security}
\scriptsize
\begin{tabular}{lp{3.5cm}rr}
\toprule
\textbf{Injected Scenario} & \textbf{Observed Broker Behavior} & \textbf{Rejection Rate} & \textbf{Unsafe Escape Rate} \\
\midrule
Uncertified Mutation & Rejected by signature verifier & 100.0\% & 0.0\% \\
Request-Cert Mismatch & Rejected by contract match & 100.0\% & 0.0\% \\
Replayed $(cid, nonce)$ & Rejected by DB nonce constraint & 100.0\% & 0.0\% \\
Stale Policy Epoch & Rejected by epoch check & 100.0\% & 0.0\% \\
Stale Revocation Epoch & Rejected by revocation check & 100.0\% & 0.0\% \\
Revocation Partition & Failed closed after cache expiry & 100.0\% & 0.0\% \\
Malformed $\Omega$ & Rejected by JSON parser & 100.0\% & 0.0\% \\
Live-State Subnet Drift & Rejected by drift predicate & 100.0\% & 0.0\% \\
Parameter-Level Violation & Rejected by proxy validation & 100.0\% & 0.0\% \\
\bottomrule
\end{tabular}
\end{table*}

In all cases, the broker successfully blocked unauthorized mutations, consistent with the enforcement predicates under the stated deployment assumptions of Theorem 1. When the Revocation Service was partitioned, the broker successfully failed closed immediately after the local cache TTL (5 seconds) expired, rejecting 100.0\% of requests.

\textbf{Broker Crash Recovery:} We simulated broker crashes during execution. Out of 50 crash events, 25 were injected prior to target API invocation. All 25 failed closed, with no mutations occurring. The remaining 25 were injected during or immediately after target API invocation. Using idempotency tokens, the recovery manager resolved the state of each pending transaction against the target cloud APIs, recording 100.0\% of outcomes correctly in the ledger without duplicate mutations. Where the target platform or API lacks reliable idempotency or lookup mechanisms, \seb conservatively treats the certificate as consumed and requires re-admission through \sab.

\subsection{Revocation Propagation Delay (Exp 3)}
To measure revocation propagation delay (RQ2), we issued certificates and queued them for execution. We then advanced the global revocation epoch at the Revocation Service and measured the time until the broker rejected execution requests.
With a polling interval of 5 seconds and a cache TTL of 5 seconds, the maximum observed delay from epoch advancement at the revocation service to 100\% request rejection at the broker was 5.2 seconds (mean delay of 2.6 seconds). This propagation delay is bounded by the polling interval plus network propagation latency.

\subsection{Live-State Drift Sensitivity (Exp 4)}
We evaluated the drift-check predicate $\Phi_{drift}$ by modifying target resources in the background during the interval between certificate signing ($t_1$) and execution request ($t_2$).
Under simulated configuration updates (e.g., changing security group descriptions or tags), the broker successfully detected state deviations. When the drift tolerance threshold $\epsilon_C$ was set to 0 (strict matching), the broker rejected 100\% of requests that experienced any background modification, showing a drift rejection rate of 4.2\% in our concurrent workload simulator.

\subsection{Performance Analysis (Latency and Throughput)}
We analyzed the latency distribution and throughput scalability under increasing request concurrency.

\textbf{Latency Percentiles:} For AWS workloads, the latency percentiles are p50 = 136.9 ms, p95 = 180.4 ms, and p99 = 284.1 ms. The tail latency is dominated by AWS STS API latency spikes. For Kubernetes workloads, the percentiles are p50 = 28.2 ms, p95 = 36.5 ms, and p99 = 47.1 ms, demonstrating a highly stable profile.

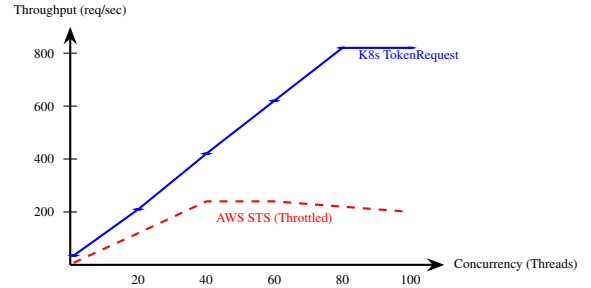
\begin{figure}[t]
  \centering
  \begin{tikzpicture}[xscale=0.45, yscale=0.035]
    \draw [thick, -Stealth] (0,0) -- (11,0) node [right, font=\tiny] {Concurrency (Threads)};
    \draw [thick, -Stealth] (0,0) -- (0,90) node [above, font=\tiny] {Throughput (req/sec)};
    
    \foreach \x/\label in {2/20, 4/40, 6/60, 8/80, 10/100}
      \draw (\x, 0.2) -- (\x, -0.2) node [below, font=\tiny] {\label};
      
    \foreach \y/\label in {20/200, 40/400, 60/600, 80/800}
      \draw (0.2, \y) -- (-0.2, \y) node [left, font=\tiny] {\label};
      
    \draw [thick, blue, mark=*] plot coordinates {(0.1,3.5) (2,21) (4,42) (6,62) (8,82) (10,82)};
    \node [blue, right, font=\tiny] at (8.2, 79) {K8s TokenRequest};
    
    \draw [thick, red, dashed, mark=square*] plot coordinates {(0.1,0.8) (2,12) (4,24) (6,24) (8,22) (10,20)};
    \node [red, below, font=\tiny] at (6, 23) {AWS STS (Throttled)};
  \end{tikzpicture}
  \caption{Throughput scalability of the \seb prototype service.}
  \label{fig:throughput}
\end{figure}

\textbf{Throughput Scalability:} Figure~\ref{fig:throughput} shows the throughput of the broker service under increasing concurrent threads.
The Kubernetes TokenRequest adapter scales linearly, peaking at 820 requests per second at 80 threads, limited primarily by PostgreSQL transaction commit rates.
Conversely, the AWS STS adapter peaks at 240 requests per second at 40 threads, beyond which AWS STS rejects requests with \texttt{RequestLimitExceeded} errors. This highlights cloud provider API throttling as a dominant operational bottleneck for broker deployments.

\subsection{Workload Suitability Discussion}
Our results show that \seb is highly suitable for high-risk, low-frequency mutation workloads. For actions like firewall rule changes, auto-scaling modifications, or database exports, the ~140 ms latency overhead is negligible relative to the security assurance of complete mediation, zero standing privileges, and cryptographic non-repudiation.
However, \seb is not suitable for high-frequency, microsecond-scale control loops (e.g., real-time container CPU scheduling or fast SDN routing updates). In those contexts, the overhead of database writes, cryptographic verification, and cloud provider API calls exceeds the microsecond-level latency budgets, necessitating alternative lightweight assurance patterns.

%% file: sections/table-metrics.tex
\begin{table}[t]
\centering
\caption{Metrics used in the empirical evaluation of the \seb prototype.}
\label{tab:metrics}
\small
\begin{tabular}{lp{3.8cm}}
\toprule
\textbf{Metric} & \textbf{Meaning / Operational Description} \\
\midrule
Broker verification latency & Latency added by signature, epoch, and contract checks \\
Scoped credential latency & Time required to mint transient tokens from IAM/STS \\
Revocation propagation delay & Elapsed time from epoch advancement to request rejection \\
Drift rejection rate & Percentage of mutations rejected due to live-state drift \\
Unsafe execution escape & Rate of invalid or uncertified mutations that succeed \\
Throughput & Number of concurrent certified mutations executed per second \\
Ledger overhead & Disk storage write overhead in bytes per transaction \\
\bottomrule
\end{tabular}
\end{table}

%% file: sections/09-security-analysis.tex
\section{Security Analysis}
\label{sec:security-analysis}

SEB's security claim is certificate-bound mutation under explicit trust
and deployment assumptions.

\subsection{Security Properties}
The Sovereign Execution Broker enforces the following security properties
when deployed as the mandatory mutation path described in
Section~\ref{sec:bypass-pattern}:
\begin{itemize}
    \item \textbf{Property 1 (No Direct Agent Mutation - P1):} If target infrastructure APIs accept mutations only from authorized mutation principals, then an autonomous agent cannot execute direct mutations on production state.
    \item \textbf{Property 2 (Certificate-Bound Execution - P2):} A mutation operation $op$ on resource $target$ executes through \seb only after it is bound to a cryptographically valid SAB certificate $\Omega$.
    \item \textbf{Property 3 (Request-Contract Matching - P3):} A certificate $\Omega$ containing contract $C$ cannot be used to authorize an execution request $req$ unless $req$ matches $C$ exactly or satisfies the explicit constraints in $C$.
    \item \textbf{Property 4 (Revocation-before-Execution - P4):} A certificate $\Omega$ can be invalidated after admission by \sab but prior to its execution by \seb by incrementing the system's global revocation epoch $\rho_{active}$.
    \item \textbf{Property 5 (Drift-Aware Execution - P5):} A certificate $\Omega$ issued under stale system evidence cannot execute if the live target state $S_t$ has drifted beyond the contract-defined safety threshold $\epsilon_C$.
    \item \textbf{Property 6 (Scopeable Identity - P6):} The broker mints transient credentials only when $Scopeable(C, Platform)$ holds; native IAM/RBAC enforce action, resource, and time constraints where supported, and broker-side proxy or admission controls enforce remaining parameter-level constraints.
    \item \textbf{Property 7 (Replay Resistance - P7):} A certificate nonce can be reserved only once, so a previously used $(cid, nonce)$ pair cannot authorize a second mutation.
    \item \textbf{Property 8 (Decision and Outcome Evidence - P8):} Verification failures produce signed decision records $D$, while successful verification followed by target invocation produces signed outcome records $O$, preventing deniability about broker decisions and mutation attempts.
\end{itemize}

\subsection{Assumptions for Theorem 1}
The theorem below assumes the following trusted components and deployment
conditions:
\begin{itemize}
    \item No agent runtime or wrapper has standing credentials that authorize production mutation.
    \item Target APIs accept production mutations only from authorized mutation principals.
    \item The \seb runtime, \sab signing key, revocation service, broker ledger, and target IAM/RBAC enforcement mechanisms are trusted.
    \item Every scope dimension claimed by $Scopeable(C, Platform)$ is enforceable by the target platform or by a mandatory broker-controlled proxy/admission path.
    \item Break-glass identities are outside the autonomous agent path and do not share credentials with agent runtimes.
\end{itemize}

\subsection{Safety Proof Sketch}
We present a formal claim regarding the security properties of the broker
enforcement model.

\paragraph{Theorem 1 (Certificate-Bound Mutation).}
\emph{Under the assumptions above, an untrusted agent runtime cannot
cause a production state mutation on target infrastructure unless the
requested mutation matches a valid, unexpired, unrevoked, unreplayed
certificate issued by the admission layer, live-state drift remains
within the admitted contract bounds, and the required scope is
enforceable by the target platform or by a mandatory broker-controlled
proxy/admission path.}

\paragraph{Proof Sketch.}
We prove the theorem by contradiction. Suppose an agent successfully
causes a state mutation $M$ at time $t_{exec}$ on target resource $R$
using operation $op$ with parameters $params$, but at least one theorem
condition is false.

\paragraph{Case 1: Broker Bypass.}
The agent attempts to execute $M$ directly without \seb. By assumption,
target APIs reject all mutation identities outside the authorized
mutation-principal set. The agent has no standing mutation credentials.
Therefore, the target API rejects the direct request, contradicting the
assumption that the agent caused $M$.

\paragraph{Case 2: No Valid Certificate or Replay.}
The agent submits a request to \seb without a valid certificate or with a
previously used nonce. The broker evaluates $\Phi_{sig}$ and
$\Phi_{replay}$:
\begin{equation}
    Verify(\Omega, \sigma_{SAB}) = \text{True}
\end{equation}
\begin{equation}
    \Phi_{replay}(cid, nonce, L) = ((cid, nonce) \notin L_{used})
\end{equation}
If the signature is invalid, the broker writes a reject decision record.
If the nonce has already been reserved, the broker writes a replay
decision record. In both cases, no new execution identity is minted,
contradicting the assumption that the broker-mediated request caused
$M$.

\paragraph{Case 3: Certificate is Expired or Revoked.}
Assume a certificate $\Omega$ exists but is outside $T_{valid}$ or has
been revoked prior to execution, meaning $\rho_{rev} < \rho_{active}$.
When the broker receives the execution request, it evaluates
$\Phi_{time}$ and $\Phi_{rev}$:
\begin{equation}
    \Phi_{rev} = (\rho_{rev} = \rho_{active})
\end{equation}
The failed predicate causes \seb to write a decision record, reject the
request, and refuse to mint $ID_{exec}$. Thus, the agent cannot execute
$M$ through the broker. Contradiction.

\paragraph{Case 4: Request-Contract Mismatch or Unsafe Drift.}
Assume a valid certificate $\Omega$ exists for contract $C$, but the
agent attempts to execute a different operation, resource, or parameter
set, or the live state has drifted beyond $\epsilon_C$. The broker
evaluates:
\begin{equation}
\begin{split}
    \Phi_{match} ={} & (req.cid = \Omega.cid) \land \\
    & (req.nonce = \Omega.nonce) \land \\
    & Match(req.op, req.target, req.params, C)
\end{split}
\end{equation}
and $\Phi_{drift}(E_{admit}, S_t)$. If either predicate is false, the
broker writes a decision record, rejects the request, and does not mint
$ID_{exec}$. Therefore, the agent cannot cause $M$ through \seb.
Contradiction.

\paragraph{Case 5: Required Scope Cannot Be Enforced.}
Assume a scope dimension required by $C$ is not enforceable by native
target-platform controls or by a mandatory broker-controlled
proxy/admission path. Then $Scopeable(C, Platform)$ is false, so the
broker writes a decision record and rejects before credential minting.
Conversely, if $Scopeable$ is true, the trusted target IAM/RBAC or
mandatory broker-controlled proxy/admission path enforces the claimed
action, resource, time, and parameter constraints. Thus, the agent cannot
use a broker-minted identity to execute outside $C$. Contradiction.

Therefore, under the stated assumptions, every successful production
mutation attributable to an untrusted agent must pass broker verification
against a valid, unexpired, unrevoked, unreplayed certificate whose
contract matches the request and whose required scope is enforceable by
the target platform or by a mandatory broker-controlled proxy/admission
path. This proof does not claim semantic safety of the admitted proposal,
which remains the responsibility of \sab and its validators; it claims
enforcement of certificate-bound authority at mutation time.
$\hfill\blacksquare$

%% file: sections/10-discussion-limitations.tex
\section{Discussion and Limitations}
\label{sec:discussion}

Operational deployment of \seb involves trade-offs around human escalation paths, performance, scaling, and target-platform expressiveness.

\subsection{Human-in-the-Loop Escalation}
While \seb enforces automated fail-closed behavior, production systems require mechanisms to handle unexpected failures, such as networking outages or partition of the revocation service. Break-glass operation is outside the autonomous certificate-bound theorem in Section~\ref{sec:security-analysis}. It is a separate human-authorized emergency path with independent authentication, authorization, and audit requirements:
\begin{enumerate}
    \item \textbf{Break-Glass Override:} Certified human operators can sign an emergency override request $\Omega_{override}$ under a policy separate from autonomous SAB certificates. 
    \item \textbf{Verifiable Escalation:} The broker verifies the operator's cryptographic signature against administrative keys and executes only if the emergency policy authorizes the mutation.
    \item \textbf{Audit Trail:} Every manual override generates an alert and is permanently written to the Ledger with high priority.
\end{enumerate}
The presence of this path does not weaken Theorem~1 because break-glass credentials are not available to agent runtimes and are not treated as certificate-bound autonomous mutations.

\subsection{Performance Scaling and Consensus}
Integrating a broker into every mutation path introduces latency. The primary bottlenecks are:
\begin{itemize}
    \item \textbf{Drift Checks:} Querying live infrastructure state requires making API calls to cloud providers, which can take hundreds of milliseconds.
    \item \textbf{Credential Minting:} Calling STS or service account APIs to generate transient tokens adds round-trip networking overhead.
\end{itemize}
To mitigate these, a deployment can use concurrent queries and cached state projections where policy permits. However, maintaining strong consistency of the revocation epoch $\rho_{active}$ across multiple distributed broker nodes requires consensus. A deployment can use a lightweight replicated epoch store, such as a Raft-backed database, to ensure all broker nodes validate requests against the same epoch number.

\subsection{Limitations}
Our design has several limitations:

\paragraph{Trust in Cloud Providers.} \seb relies on the target cloud provider's IAM system to enforce the boundaries of the minted scoped token $ID_{exec}$. If the provider's IAM enforcement layer has vulnerability bugs or allows bypass, the security properties in Section~\ref{sec:security-analysis} no longer hold.

\paragraph{Cold-Start Latency.} Minting fresh session tokens dynamically for every single mutation request prevents credential reuse, but suffers from cold-start latencies. For high-throughput microsecond-level mutation workloads, the overhead of constant credential generation can become prohibitive, necessitating token caching strategies that introduce security trade-offs.

\paragraph{Target API Limitations.} Not all infrastructure APIs support granular parameter scoping. For example, some cloud APIs allow scoping by resource tags but do not permit scoping the specific JSON payload parameters. In these cases, the deployment must use proxy-based parameter validation to make $Scopeable(C, Platform)$ true, which increases complexity and introduces a potential point of failure.

\paragraph{Broker Availability as a Critical Dependency.} Since the broker acts as the mandatory mutation gatekeeper, any outage of the broker blocks all agent mutations, representing a single point of failure (SPoF). Maintaining high availability (HA) with multi-AZ replication, database failover, and rate limiting is required, adding substantial infrastructure overhead.

\paragraph{Operational Complexity.} Transitioning legacy environments to a zero-standing-privilege posture is operationally complex. It requires configuring and maintaining Service Control Policies (SCPs), IAM permission boundaries, validating admission webhooks, and reverse proxy paths across all active mutation vectors.

\paragraph{Revocation Consistency vs. Load.} Polling the global revocation epoch at a short interval (e.g., 5 seconds) minimizes the vulnerability window for revoked certificates. However, in large deployments with thousands of broker instances, frequent polling generates significant read load on the central revocation service, requiring caching strategies that introduce consistency delays.

\paragraph{State-Observation Staleness.} Cloud provider APIs are eventually consistent. During drift checks, the broker queries the target state; however, if the cloud control plane has not fully propagated a background change, the broker may evaluate drift based on stale observations, introducing time-of-check to time-of-use (TOCTOU) windows.

\paragraph{Semantic Safety Remains External.} The broker is an execution-time enforcement boundary. It guarantees that the request matches the certificate, but it cannot verify if the certificate's contract itself is semantically safe. If SAB (or SQA) admits a faulty or malicious proposal due to validator compromise, SEB will execute it.

\paragraph{Cross-Cloud and Multi-Account Deployments.} In heterogeneous multi-cloud environments (spanning AWS, Azure, Google Cloud, and Kubernetes), maintaining unified trust roots, global revocation epochs, and consistent proxy layers is challenging due to the differences in platform credential-minting and policy mechanisms.

%% file: sections/11-related-work.tex
\section{Related Work}
\label{sec:related-work}

SEB bridges ideas from capability systems, reference monitors, privileged access management, workload identity, policy-as-code, and safety-critical shields, adapting them to the unique requirements of agentic control planes.

\subsection{Reference Monitors and Complete Mediation}
Reference monitors mediate all access requests to protected resources, enforcing properties such as complete mediation, tamper-proofing, and verifiability~\cite{anderson-reference-monitor, schneider-enforceable}. 
\begin{itemize}
    \item \textbf{What SEB borrows:} SEB adopts the structural design of a classical reference monitor, interposing between the client (the agent wrapper) and the resource (the target infrastructure APIs).
    \item \textbf{What SEB adds:} Traditional reference monitors operate at the operating system or kernel level (mediating system calls). SEB implements complete mediation at the cloud control-plane and API layers. It achieves this by enforcing a zero-standing-privilege network topology where all mutation paths reject non-broker credentials, and by binding execution to dynamic admission certificates rather than static ACLs.
\end{itemize}

\subsection{Privileged Access Management and Zero Standing Privilege}
Zero Standing Privilege (ZSP) and Just-in-Time (JIT) access aim to reduce attack surfaces by issuing temporary administrative credentials only when an active task requires them~\cite{rose-zerotrust, aws-iam-roles-anywhere}.
\begin{itemize}
    \item \textbf{What SEB borrows:} SEB builds on the ZSP philosophy, denying agents standing write credentials and issuing short-lived, transient credentials.
    \item \textbf{What SEB adds:} Traditional PAM systems are human-centric, validating developer identities, MFA tokens, or approval tickets to open a time-bounded session. SEB automates this for non-deterministic agents, mapping credential scope directly to a cryptographically signed *action contract* ($C$) rather than a principal's identity. Credential authority is bound to a single-use transaction.
\end{itemize}

\subsection{Cloud Control-Plane Authorization and Workload Identity}
Workload identity systems (e.g., SPIFFE/SPIRE) issue cryptographic service identities to workloads~\cite{spiffe, spire}, while cloud platforms provide mechanisms like AWS STS session policies and Kubernetes projected tokens to down-scope permissions~\cite{aws-iam, aws-sts-session-policy}.
\begin{itemize}
    \item \textbf{What SEB borrows:} SEB utilizes workload identity for broker service authentication and leverages STS/TokenRequest APIs to generate down-scoped credentials.
    \item \textbf{What SEB adds:} Workload identity verifies *who* is running but does not evaluate *what* the workload intends to do or whether it is safe. SEB acts as a controller layer on top of these identity substrates. It dynamically translates the signed, high-level assurance contract $C$ from SAB into corresponding JSON-formatted session policies, session tags, and projected tokens, while interposing a validation proxy to enforce parameter-level constraints that the underlying cloud IAM cannot natively express.
\end{itemize}

\subsection{Policy-as-Code and Admission Control}
Policy decision engines such as Open Policy Agent (OPA), Gatekeeper, Kyverno, and Cedar decouple authorization logic from target application platforms~\cite{opa-rego, gatekeeper, kyverno, cedar}.
\begin{itemize}
    \item \textbf{What SEB borrows:} SEB leverages the declarative payload inspection concepts popularized by policy-as-code engines.
    \item \textbf{What SEB adds:} Policy-as-code engines are stateless Policy Decision Points (PDPs) that evaluate static rules against incoming requests. They do not handle credential-minting custody, database nonce reservations, or target-state drift checks. SEB is a stateful Policy Enforcement Point (PEP) and reference monitor that integrates policy validation with replay protection, cache checks, and signed ledger logging.
\end{itemize}

\subsection{Runtime Enforcement for Safety-Critical Systems}
Safety shields and Simplex-style architectures interpose a safety controller to override or block unsafe commands from a primary controller in physical systems~\cite{simplex-architecture, bloem-shield-synthesis, alshiekh-shielding}.
\begin{itemize}
    \item \textbf{What SEB borrows:} SEB borrows the shield concept, interposing a validator to ensure commands satisfy predefined safety boundaries.
    \item \textbf{What SEB adds:} Simplex shields focus on continuous reactive feedback loops in physical environments (e.g., flight controls). SEB extends this to discrete cloud infrastructure mutations, translating abstract safety boundaries into ephemeral IAM policies and validating them against validity windows, revocation epochs, and database transaction logs.
\end{itemize}

\subsection{Supply-Chain Attestations and Provenance}
Software supply-chain frameworks (e.g., in-toto, SLSA) bind build steps and materials to cryptographic attestations of provenance~\cite{in-toto, slsa}. Replicated consensus protocols (e.g., Paxos, Raft) are used to order events durably~\cite{lamport-paxos, ongaro-raft}.
\begin{itemize}
    \item \textbf{What SEB borrows:} SEB borrows the concept of cryptographically signed build/execution provenance and utilizes append-only log structures.
    \item \textbf{What SEB adds:} Supply-chain systems verify provenance post-hoc (after the build completes). SEB evaluates provenance \emph{pre-execution} (asserting that a mutation matches the SAB certificate) and binds the \emph{post-execution} outcome. By signing both \textsc{DecisionRecord}s and \textsc{OutcomeRecord}s, SEB builds an active, tamper-proof runtime audit trail that connects the agent's intent to the final system state.
\end{itemize}

%% file: sections/12-conclusion.tex
\section{Conclusion}
\label{sec:conclusion}

As non-deterministic agentic control planes assume greater operational authority over production systems, traditional authorization models based on static identity privileges become untenable. This paper introduced the Sovereign Execution Broker (\seb), a runtime enforcement boundary that addresses the gap between proposal admission and runtime infrastructure mutation. 

By separating proposal, admission, and execution, \seb removes standing mutation credentials from autonomous agents when target APIs are configured to reject non-broker identities, and turns certified authority into short-lived, revocable, and auditable runtime capabilities. We presented the \seb execution model, formalized its certificate, nonce, record, and scopeability semantics, and detailed its scoped identity and fail-closed behaviors. 

We implemented a Go-based prototype of \seb and evaluated its performance on AWS and Kubernetes. Our evaluation shows that \seb adds manageable latency overheads of approximately 28 ms for Kubernetes mutations and 195 ms for AWS mutations, dominated by cloud API calls and credential minting, scales to hundreds of requests per second, and successfully rejects stale, replayed, or uncertified requests under injected faults. SAB certifies proposal authority, but \seb is the mandatory runtime enforcement boundary that turns certificates into fresh, scoped, revocable mutation authority.